\documentclass[aps,twocolumn,prb,superscriptaddress,floatfix,showpacs]{revtex4}
\usepackage{amsmath}
\usepackage{graphicx}
\usepackage{subfigure}
\usepackage{color}
\usepackage{epstopdf}

\def\lsim{\mathrel{\mathpalette\gl@align<}}
\def\gsim{\mathrel{\mathpalette\gl@align>}}
\def\gl@align#1#2{\lower.6ex\vbox
{\baselineskip\z@skip\lineskip\z@
\ialign{$\m@th#1\hfil##\hfil$\crcr#2\crcr\sim\crcr}}}

\makeatother

\newcommand\ba{\begin{eqnarray}}
\newcommand\ea{\end{eqnarray}}
\newcommand\be{\begin{equation}}
\newcommand\ee{\end{equation}}
\newcommand\bi{\bibitem}
\newcommand{\ct}{\cite}

\begin{document}

\title{Slow quenches in a quantum Ising chain; dynamical phase transitions and topology}

\author{Shraddha Sharma}
\affiliation{Department of Physics, Indian Institute of
Technology, Kanpur 208 016, India}
\author{Uma Divakaran}
\affiliation{UM-DAE Center for Excellence in Basic Sciences, Mumbai 400 098, India}
\author{Anatoli Polkovnikov}
\affiliation{Department of Physics, Boston University, 590 Commonwealth Ave., Boston, MA 02215, USA}
\author{Amit Dutta}
\affiliation{Department of Physics, Indian Institute of
Technology, Kanpur 208 016, India}


\begin{abstract}
 We study the slow quenching dynamics (characterized by an inverse rate, $\tau^{-1}$) of a one-dimensional transverse Ising chain with nearest neighbor ferromagentic interactions across the quantum critical point (QCP) and  analyze  the Loschmidt overlap  {measured using the subsequent temporal evolution of the  final wave function (reached at the end 
 of the quenching)  with the final time-independent Hamiltonian}. 
  Studying the Fisher zeros of the corresponding generalized ``partition function", we probe  non-analyticities
 manifested in the rate function of the return probability known as dynamical phase transitions (DPTs). In contrast  to the sudden quenching case, we show that DPTs survive {in the subsequent   temporal evolution following the quenching across
 two critical points of the model for a sufficiently slow
 rate; furthermore, an interesting ``lobe" structure of Fisher zeros emerge.} We have also made a connection to topological aspects studying the dynamical topological order parameter ($\nu_D(t)$),
 as a function of time ($t$) {measured from the instant when the quenching is complete. Remarkably, the time evolution of $\nu_D(t)$ 
exhibits drastically different behavior following quenches across a single QCP and two QCPs. }  {In the former case, $\nu_D (t)$ increases 
step-wise by unity at every DPT (i.e., $\Delta \nu_D =1$). In the latter case, on the other hand, $\nu_D(t)$ essentially oscillates between 0 and 1 (i.e., successive DPTs occur with $\Delta \nu_D =1$ and $\Delta \nu_D =-1$, respectively), except for instants where it
 shows a sudden jump by a factor of unity when two successive DPTs carry a topological charge of same sign.}
 
\end{abstract}


\maketitle



\section{Introduction}

The phase transition in a thermodynamic system is marked by  non-analyticities in the free-energy density which
can be detected by  analyzing the zeroes of the partition function in a complex temperature plane as proposed by Fisher \ct{fisher65}; a similar proposal was also given earlier in the presence of a complex magnetic field \cite{lee52} (See also [\onlinecite{saarloos84}]). When   the line of Fisher zeros cross the real axis, there are  non-analyticities in the free-energy density which signal
the existence of a finite temperature phase transition in the thermodynamic limit.

 In a recent work, Heyl {\it et al.} \ct{heyl13},
introduced the notion of {\it dynamical phase transitions} (DPTs) exploiting the formal similarity between the canonical partition function
$Z(\beta)= \textup{Tr}~ e^{-\beta H}$,
 of an equilibrium system described by a Hamiltonian $H$ (where $\beta$ is the inverse temperature) and that of the overlap amplitude or Loschmidt overlap (LO) defined for a quantum system which is suddenly quenched. Denoting  the ground state of the initial Hamiltonian as  $|\psi_0\rangle$ and the final Hamiltonian reached through the
quenching process as  $H$, the LO is defined as  $G(t)=\langle\psi_0|e^{-iHt}|\psi_0\rangle$. Generalizing $G(t)$  to 
 $G(z)$ defined in the  complex  time ($z$) plane, one can introduce the notion of a dynamical free energy density, $f(z)=-\lim_{L\to \infty} \ln{G(z)}/L^d$, where $L$ is the linear dimension of a $d$-dimensional system. 

 In a spirit  similar to the classical case,
one  then looks  for the zeros of the $G(z)$ (or non-analyticities in $f(z)$) to define a dynamical phase transition.  For a transverse Ising chain, it has been observed \ct{heyl13}  that  when the system is suddenly quenched across the quantum critical point (QCP) \ct{Sachdev,suzuki13}, the line of  Fisher zeros  crosses the imaginary time axis  at  instants $t^{*}$; at these instants  {the rate function of the return probability} defined as $I(t) = - \ln |G(t)|^2/L$ shows sharp non-analyticities.    

The initial observation by  Heyl {\it{et al.}}
\ct{heyl13}  was verified in several subsequent studies \ct{karrasch13,kriel14,andraschko14,canovi14,heyl15,palami15}  which established that similar DPTs are observed for  sudden quenches across the QCP for both integrable and non-integrable models. Subsequent  works  however showed that DPTs can
occur  following a sudden quench  even within the same phase (i.e., not crossing the QCP) for both integrable \ct{vajna14} as well as non-integrable models \ct{sharma15}.
Furthermore, DPTs have been explored in  two-dimensional systems \ct{vajna15,schmitt15} where topology of
the equilibrium system 
plays a non-trivial role \ct{vajna15};  also, the notion of a local dynamical topological order parameter (DTOP) which assumes
integer values and   changes
by unity at every DPT  has been proposed \ct{budich15}. 
We note in the passing that the rate function $I(t)$  is related to the Loschmidt echo which has been
studied in the context of decoherence at zero   \ct{quan06,rossini07,cucchietti07,venuti10,sharma12,nag12,mukherjee12,dora13,sharma14} and also at finite temperatures \ct{zanardi07_echo} and  has also been  useful in studies   of the work-statistics \ct{gambassi11} and the entropy generation \ct{dorner12,sharma15_PRE,russomanno15} in quenched quantum systems. {In fact, the rate function (of the return probability) discussed above in the context of DPTs
can be connected to the
singularities in the work distribution function corresponding to the zero work following a double quenching experiment\ct{heyl13}; in this process, the initial Hamiltonian is suddenly quenched to the final Hamiltonian at $t=0$ and then quenched back to the initial one
at a time $t$.}

It should be noted that the periodic occurrences of non-analyticities in the rate function for an integrable model was first reported in the reference [\onlinecite{pollmann10}], in
the context of a slow quenching of the transverse Ising chain though the connection to DPTs and Fisher zeros remain unexplored. In the present work, we shall address that particular issue and  probe  the behavior
of the line of Fisher zeros following   slow  quenches (dictated by a rate $\tau^{-1}$) of integrable spin chains across their
QCPs; we shall also connect  this to point out the occurrences of such DPTs or non-analyticities  manifested in the rate function of
the return probability. 
 Furthermore, we address the question concerning the fate of these DPTs when the transverse Ising chain is
slowly driven across both the QCPs  of the model; it is noteworthy that for the sudden quenching, the passage through two critical
points  wipes out the non-analyticities in the rate function \ct{heyl13}.   We note at the outset that in
the context of the slow quenching {there are two times denoted by ${\tilde t}$ and $t$, respectively,} appearing in the discussion here: ${\tilde t}$ is related to the ramping of a parameter of the Hamiltonian {(namely, the transverse field $h$)
using the quenching protocol} ${\tilde t}/\tau$ to reach the final state  {$|\psi_f\rangle$ at a desired final value of the field $h_f$}; the other one, {denoted by $t$},  that appears in $G(t)$  describes the subsequent temporal evolution
of {the state $|\psi_f\rangle$} with the time-independent final Hamiltonian $H_f$.  Hence, once the linear-ramping (with time ${\tilde t}$) of the parameter is complete, {we set the time $t=0$, and  probe the non-analyticities in $I(t)$ as a
function of $t$ (analytically continuing it to the complex time plane). In other words, the role of the slow quenching is to  prepare
the system in a desired state; we then probe the time evolution of this state with the  final Hamiltonian as a function of $t$}.  {Secondly, we study Fisher zeros numerically using a finite system, hence they do not really form
a line, rather generate a set of closely spaced points.}

The study of slow quenching dynamics has gained importance in recent years because of the possible Kibble-Zurek (KZ) scaling \ct{Kibble1976,Zurek1985} of the defect density and the residual energy following a quench across (or to)  a QCP by tuning a 
parameter of the Hamiltonian slowly  \ct{Zurek2005,Polkovnikov2005}. This scaling has
been verified and also modified in various situations \ct{dziarmaga05,damski05,cherng06,mukherjee2007,bib:Pellegrini,sengupta2008,deng08,dutta2010,degrandi10,thakrathi12,
canovi141}.
(For reviews, see  [\onlinecite{PolkovnikovRev,dziarmaga10,dutta15}].)

The paper is organized in the following manner: in Sec \ref{sec_trans}, we discuss a transverse Ising chain
driven across its QCP by varying the transverse field following a linear protocol;  studying the behavior of Fisher
zeros we determine the location of non-analyticities in the rate function. We study both the situations when the transverse field
is quenched across a single critical point  and both the critical points of the model, as shown in Figs.~\ref{fig:DPT_slow1}
and \ref{fig:DPT_slow2}, respectively, and compare the results with
those of the sudden quenching.       {Finally, in Sec. \ref{sec_topology}, we present the rich behavior that emerges from
the study of  the  variation of the DTOP ($\nu_D(t)$) as a function of time ($t$) following a slow quench and explore
the topology of DPTs which is dictated by the topological properties of the equilibrium system.
 We investigate both the situations, when
starting from a non-topological phase, the system is quenched to a topological phase (crossing a single  QCP) or to the other non-topological phase across both the QCPs separating the topological phase and the non-topological phases; the behavior of the DTOP is significantly different in these two situations. In the former case,  $\nu_D(t)$ increases
in a step-like fashion by unity at every DPT while in the latter it oscillates between zero and unity except at the special
instants  when 
there are two successive  DPTs with the same sign of the topological charges; in that situation $\nu_D$ shows a discrete jump
of unit magnitude. }

\section{Transverse Ising chain and quenching of the transverse field}
\label{sec_trans}

In this section, we shall study  the slow dynamics of a transverse Ising chain and illustrate the flow of the line of  the Fisher zeros 
studying the temporal evolution {as a function of time $t$} following such a quench.  We shall show that the Fisher zeros cross the imaginary time axis {in the complex time plane}  for 
a particular momentum leading to non-analyticites of the rate function at the corresponding real time.
Let us first consider a ferromagnetic transverse Ising  chain described by the Hamiltonian
\be
H =- J_x\sum_i \sigma_i^x \sigma_{i+1}^x  - h \sum_i \sigma_i^z,
\label{eq_hamil_txy}
\ee
where $\sigma_i$'s are the Pauli spin matrices satisfying the standard commutation relations, $J_x$  is
the ferromagnetic nearest neighbor interactions in the $x$ directions (set equal to unity in the subsequent
discussion) and $h$ denotes the strength of the
non-commuting transverse field.
The model being translationally invariant
(and hence momentum $k$ being a good quantum number), we use Fourier transformation and employ
Jordan-Wigner transformation to map spins to spin-less fermions. Furthermore noting that the `parity' (of the number
of Jordan-Wigner fermions) is conserved, one can arrive at decoupled $2 \times 2$ Hamiltonians for each
mode $k$ in the basis $|0\rangle$ (no fermion) and $|k,-k\rangle$ (with two fermions with quasi-momenta $k$ and
$-k$, respectively) given by  \ct{dutta15}

\begin{equation}
 H_{k}= 
 \left(
 \begin{array}{cc}
-h +\cos k & -i  \sin k  \\
i \sin k  &  h - \cos k   \\
 \end{array}
 \right).
 \label{ham_2by2}
 \end{equation} 
 Analyzing the gap in the energy spectrum ($2\epsilon_k=2\sqrt{(h -\cos k)^2+\sin^2k}$) of the Hamiltonian, it can be shown that the model has two QCPs at
 $h = \pm 1$ where the gap vanishes for the momentum modes $k=0$ and $k=\pi$, respectively. These QCPs separate the ferromagnetic phase (for $|h|<1$) from the paramagnetic phase.
 
 \begin{figure}
  \subfigure[]
  {\includegraphics[width=6cm]{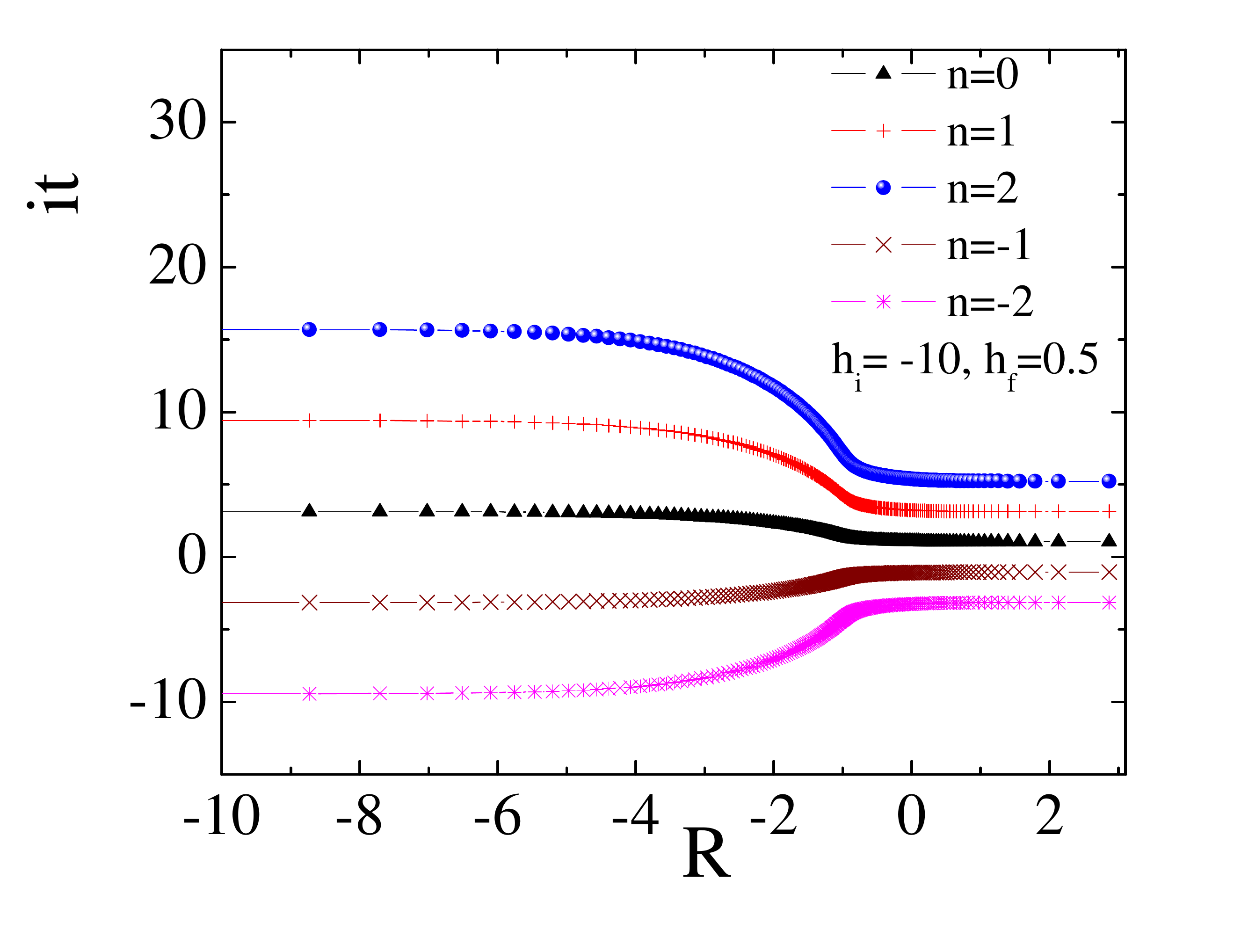}\label{fig:SS1a}}
  \subfigure[]
  {\includegraphics[width=6cm]{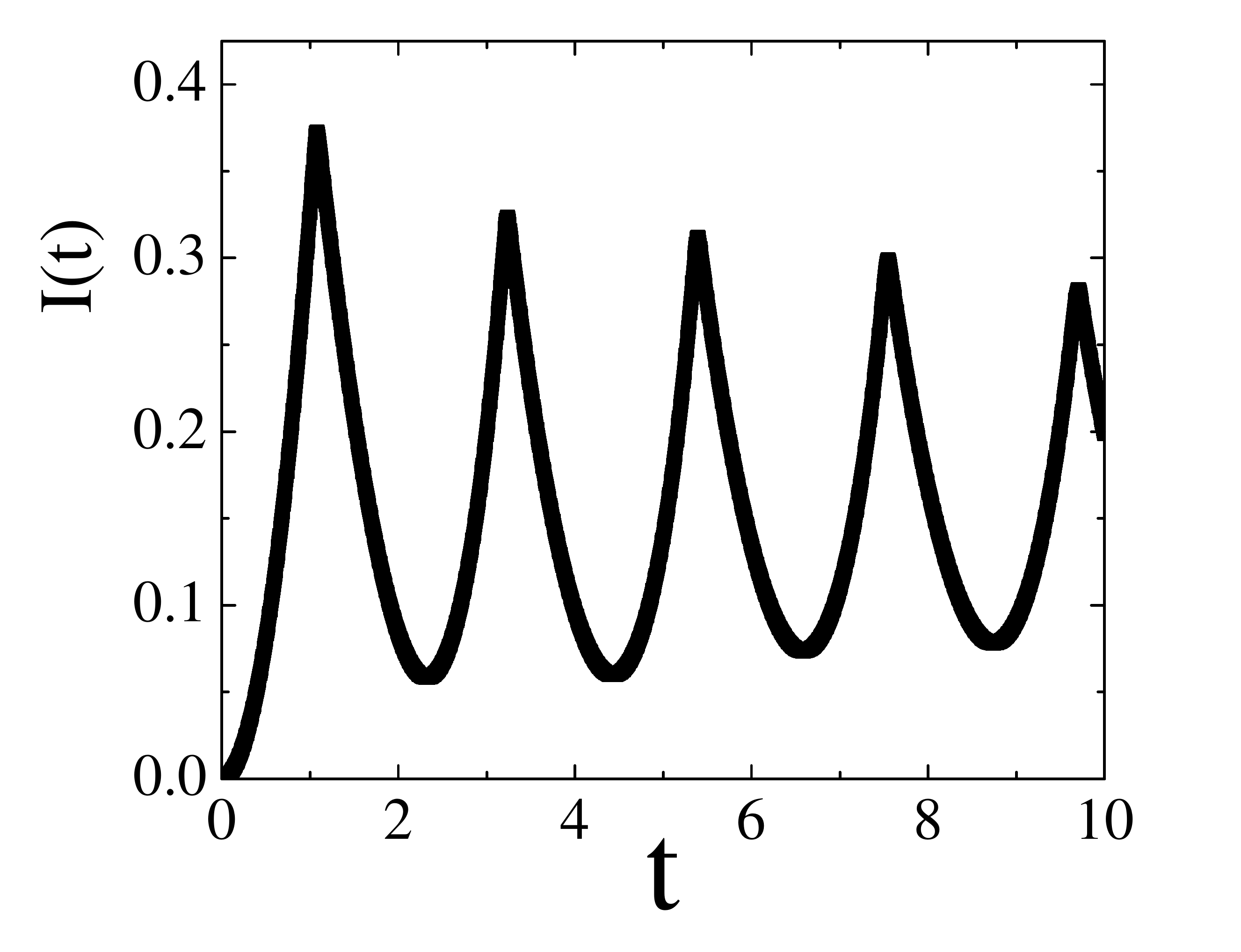}\label{fig:SS1b}}
  \caption{(Color online) (a) Fisher zeros as obtained from Eq.~(\ref{eq_fisher_zero}) are plotted in complex $z$ plane for different $n$, where $R$ denotes Re ($z$) and $it$ is the Im ($z$).  The line of Fisher zeros cross the imaginary axis for a momentum value $k_*$.  (b) Sharp non-analyticities in the rate function are observed periodically at (real) times $t_n^{*} $  (Eq.~(\ref{eq_time})) calculated at the momenta values $k_*$ as shown in the panel (a). {Here 
we have followed the quenching protocol $h({\tilde t})={\tilde t}/\tau$ with $\tau=1$, when the field is quenched from $h=-\infty$ to $h=0.5$ crossing the QCP at $h=-1$. The time $t$ is measured from the instant immediately after the quenching is complete, when we set
$t=0$.}
 }\label{fig:DPT_slow1}
\end{figure}

For all the slow quenching schemes to be analyzed in the subsequent discussions, we shall assume that the system is initially
prepared in the ground state of the initial Hamiltonian. We first consider  a linear quenching of the transverse field $h = {\tilde t}/\tau$, with ${\tilde t}$ going from a large negative value {(i.e.,  the initial field $h_i$ is large and negative)}   to a desired final value so that  $h_f <1$ (chosen to be equal to $0.5$ here); hence,  the system crosses
the QCP at $h=-1$ where the relaxation time diverges leading to the breakdown of the adiabatic evolution and hence the
final state ($|\psi_f\rangle$) reached is not the ground state of the final Hamiltonian ($H_f$).  Considering the reduced Hamiltonian (\ref{ham_2by2}), one can define the dynamical free energy \cite{heyl13}:  {$f(z) = - \ln \langle \psi_f |\exp(-H_f z)| \psi_f \rangle/L$, where $\langle \psi_f |\exp(-H_f z)| \psi_f \rangle$ is the Loscmidt overlap
while $z$ is the
complex time}. {It is to be noted that we have set $\hbar=1$ throughout.}

{Let us first illustrate the slow quenching scheme following the protocol $h={\tilde t}/\tau$, with the initial state is  $|k,-k\rangle$
for very large negative  $h$. {(In this limit, the reduced $2\times 2$ Hamiltonian in Eq. \eqref{ham_2by2} effectively becomes diagonal and hence the 
ground state, for the mode $k$, becomes $|k,-k\rangle$; on the other hand, for large positive $h$, the ground state is $|0\rangle$. Other than these
extreme cases, the ground state  is  in general  a linear superposition of these two basis states.) }}
The final state reached after the ramping (for the $k$-th mode) can then be written as  $|\psi_{f_k} \rangle = v_k |1_k^f\rangle  + u_k |2_k^f\rangle$, with $|u_k|^2 + |v_k|^2 =1$: here, $|1_k^f\rangle$ and $|2_k^f\rangle$ are the ground
state and the excited states of the Hamiltonian $H_k(h_f)$ with energy eigenvalues $-\epsilon_k^f$ and $\epsilon_k^f$, respectively. Clearly $|u_k|^2=|\langle2_k^f|k,-k\rangle|^2 =p_k$ denote the non-adiabatic transition probability that the system ends up  at the excited state after  the quench.

{Once the quenching is complete, we bring in the second time $t$, set equal to zero, and study the evolution of the state $|\psi_{f_k}\rangle$  with the final Hamiltonian $H_k(h_f)$; generalizing  to the complex time plane $z$, 
the LO  for the $k$-th mode ($L_k = \langle \psi_{f_k}| \exp(- H_k(h_f)z)|\psi_{f_k} \rangle$)  is then given by $( |v_k|^2 + |u_k|^2 \exp(-2 \epsilon_k^f z))$}. {Here, we have rescaled the ground state energy of the final Hamiltonian $H_k(h_f)$ to 0, so that the excited state energy $\epsilon_f^{\rm excited}=2\epsilon_k^f$}.
  Summing over the contributions from all the momenta modes and converting
summation to the integral, in the thermodynamic limit  we obtain:
\ba
f(z) &=& -  \int_0^{\pi} \frac{dk}{2\pi} \ln \left( |v_k|^2 + |u_k|^2 \exp(-2 \epsilon_k^f z) \right) \nonumber \\
&=& - \int_0^{\pi} \frac{dk}{2\pi} \ln \left( (1-p_k) + p_k \exp(-2 \epsilon_k^f z) \right).\ea
We then immediately find the  zeros (i.e., the Fisher zeros) of the ``effective" partition  function given by:

\be
z_n(k) = \frac 1 {2 \epsilon_k^f}  \left( \ln (\frac { p_k}{1-p_k}) + i \pi(2n+1)\right),
\label{eq_fisher_zero}
\ee
where $n=0,\pm 1, \pm 2, \cdots $. 

\begin{figure}
\centering
\subfigure[]
{\includegraphics[width=6cm]{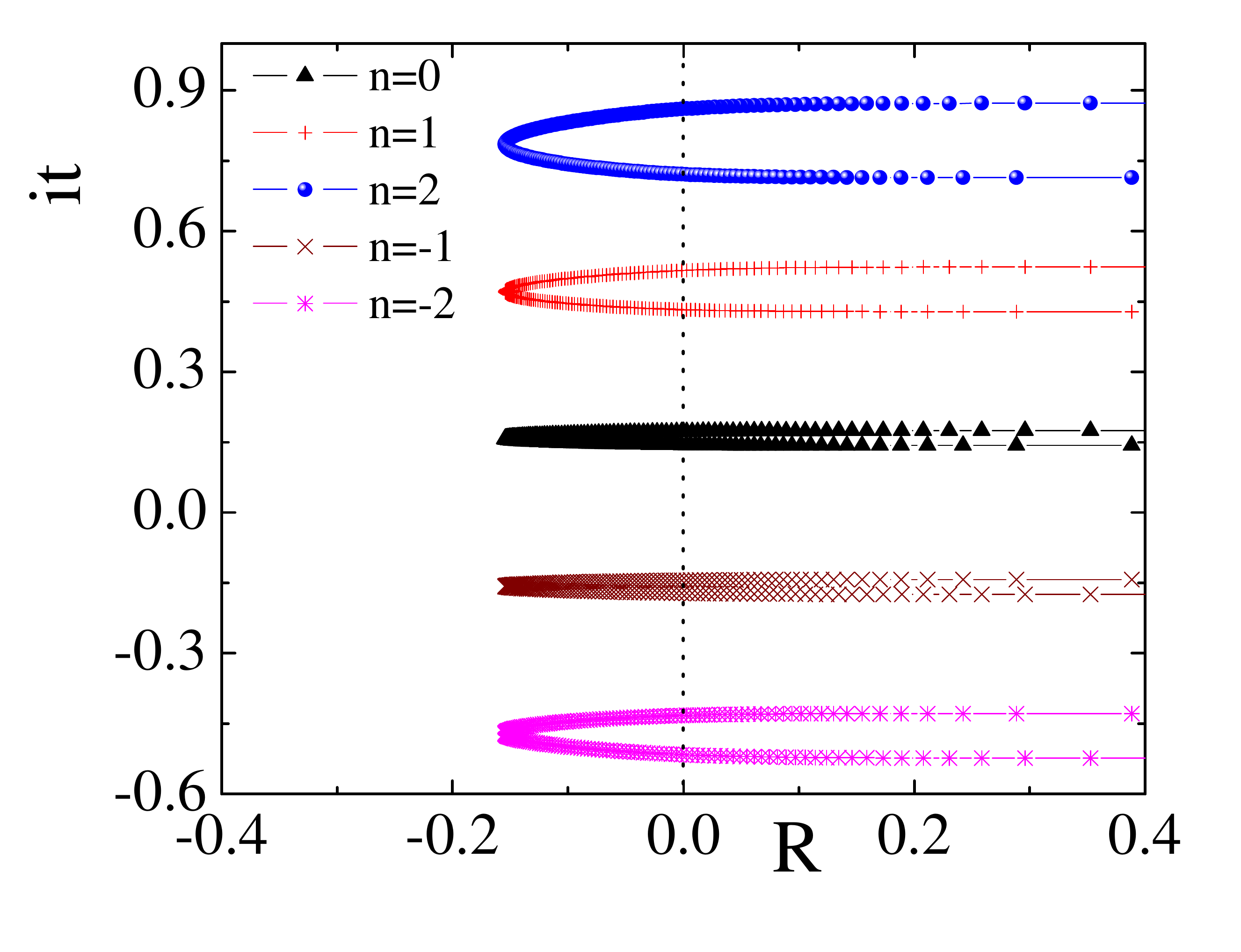}\label{fig:SS2a}}
\subfigure[]
{\includegraphics[width=6cm]{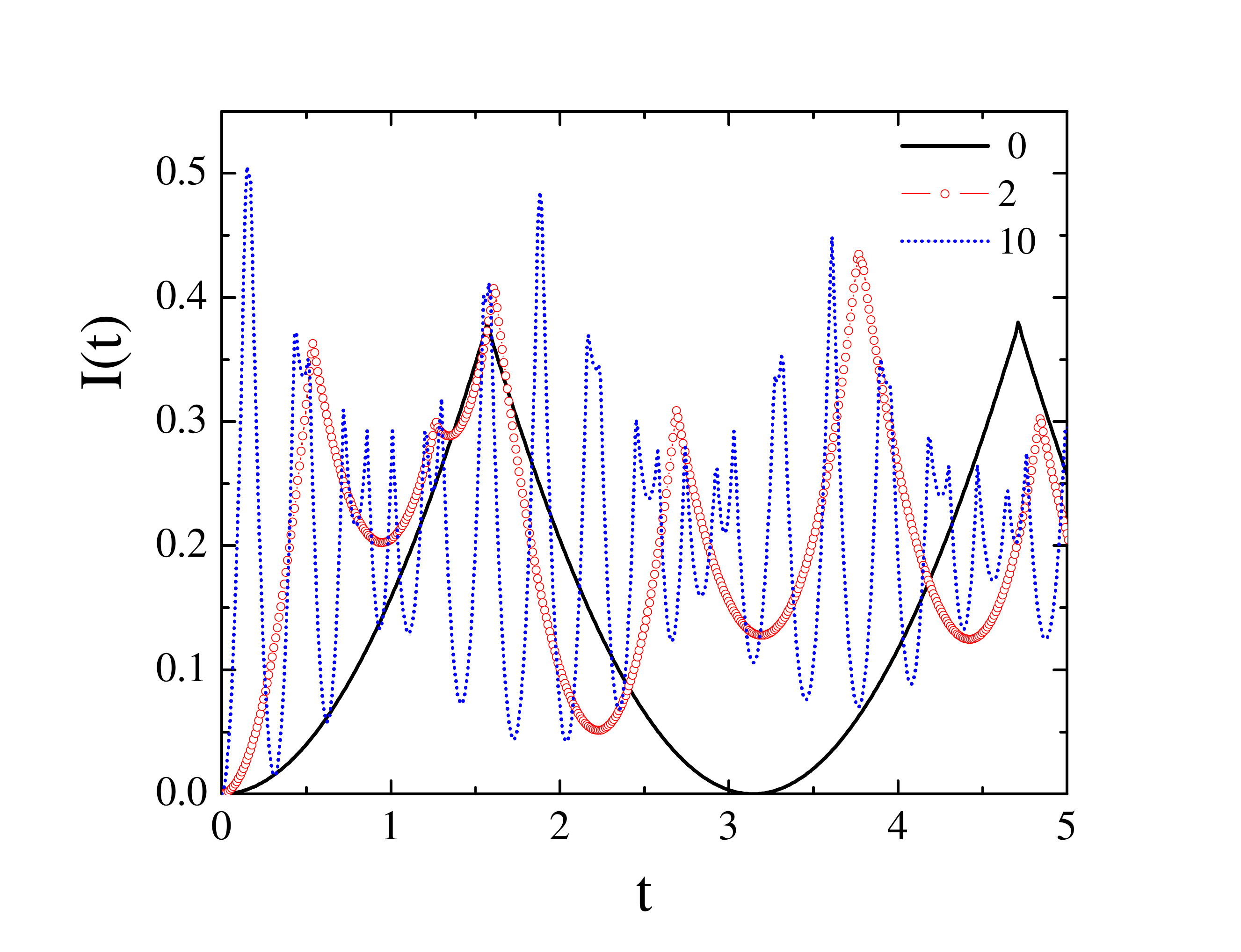}\label{fig:SS2b}}
\caption{(Color online) (a) Lines of Fisher zeros  for each integer $n$ {following quenching across two QCPs} when $h$ is quenched from $h_i=-10$ to $h_f=10$
{with $h({\tilde t })={\tilde t}/\tau$ and $\tau=1$}.       {Corresponding to each lobe denoted by $n$, we find two $k_*$s close to two different critical modes ($k=0$ and $\pi$); consequently,  there exist two $t_n^{*}$s  which we shall denote by $t_n^{+}$ and $t_n^{-}$, with $t_n^{+} <t_n^{-}$.}
(b) {The figure shows the corresponding non-analyticities in  $I(t)$ for different final values of $h$  ($h_f=0,2,10$) at different instants of times which match perfectly with those predicted by Eq. \eqref{eq_time} for $h_f=0$ and Eq.~(\ref{tns}) for $h_f=2,10$.}}


\label{fig:DPT_slow2}

 \end{figure}

The Fisher zeros following the slow quenching across one critical point and terminating the driving at $h_f=0.5$ are shown in Fig.~\ref{fig:SS1a}.
Let us now probe the question whether the line of Fisher zeros cross the imaginary axis for a particular
momenta mode $k_*$.  Clearly, this happens when $p_{k_*}=1/2$ (when the ${\rm Re} (z_n(k))$ vanishes); this in fact corresponds to ``infinite temperature" state when
both the levels of the two-level system are equally populated. Furthermore, for the critical mode $k=\pi$ (which does not evolve in the
process of quenching)
$p_k=1$ while in the case of
 high energy modes (i.e., modes close to $k \sim 0$), $p_k \to 0$ (see discussion around Eq.~\eqref{ham_2by2_small_k} and  Fig.~\ref{fig:SS_pk_half}), since according to KZ theory, modes $k > \tilde k \sim \tau^{-1/2}$ move adiabatically, i.e.,
 do not sense the passage through the QCP. Using Eq.~(\ref{eq_fisher_zero}), we therefore conclude that  $z_n (k \to 0) \to - \infty$ while for  $z_n (k \to \pi) \to + \infty$. This immediately
 implies that $z_n(k)$ indeed goes from $\infty$ to $-\infty$, crossing the imaginary time axis for a particular
 $k_*$ as shown in Fig.~\ref{fig:SS1a}. (It should be noted that  numerically  we are dealing with a finite system
 and hence Fisher zeros do not really go from $-\infty$ to $+\infty$ but they indeed cross the imaginary axis 
 for {$k_*$}.)   An exact expression for the rate function $I(t)$, (where, we reiterate, the  time $t$ is measured after the
 quenching is complete) can be exactly derived in the present case,

 \be
I(t)= - \int_{0}^{\pi} \frac{dk}{2\pi} \ln \left(1 + 4 p_k (p_k-1) \sin^2 \epsilon_k^f t \right).
\label{eq_rate_function}
\ee
As presented in Fig.~\ref{fig:SS1b}, the non-analyticities in $I(t)$ appear at the values of the real time $t_n^*$s
 
  \be 
 t_n^* = \frac {\pi} {\epsilon_{k_*}^f}  \left(n+\frac 1 {2}\right),
 \label{eq_time}
 \ee
 derived by setting ${\rm Re} (z_n(k_*))=0$ in Eq.~(\ref{eq_fisher_zero}) because the argument of $\log$ in Eq.~(\ref{eq_rate_function})
  vanishes when $k=k_{*}$ and $t=t_n^*$.
 
%

We then consider the case when the transverse field is quenched from a large negative value to a large positive value with a rate $\tau^{-1}$, crossing both the QCPs in the process; the results are presented in Fig.~\ref{fig:DPT_slow2}. The field is varied as $h({\tilde t})={\tilde t}/\tau$ so that the system crosses the critical point at  $h=-1$ at time ${\tilde t}=-\tau$ and that at $h=1$  at time ${\tilde t}=\tau$. {At the end of the ramp, the spin chain is in  the state $|\psi_{f_k} \rangle = v_k |1_k^f\rangle  + u_k |2_k^f\rangle$, (with $|u_k|^2 + |v_k|^2 =1$);  the variation of $|u_k|^2$
as a function of $k$ is shown in the  Fig.~\ref{fig:SS_pk_full}}.   
As $h({\tilde t})$ is varied the system detects both these critical points which are gapless at different $k$ values ($0$ and $\pi$). For $\tau \gg 1$,   the transition probability $p_k$ is close to unity for both $k\to 0, \pi$ and falls off exponentially for $k \gg 0$ or $k \ll \pi$.

{We now proceed to analyze the Fisher zeros and probe the non-analyticties in the rate function $I(t)$, again
setting $t=0$ when the ramping is complete;
the LO derived using the time evolution   of $|\psi_{f_k}\rangle$ with $H_k(h_f)$ generalizing the real $t$ to the complex $z$ plane. As shown
in Fig.~\ref{fig:SS_pk_full}}, the profile of $p_k$ clearly marks the fact that one is expected to obtain two separate $k _*$'s owing to these two individual critical points which can be treated  independent of each other.  Hence, although  $z_n(k) \to \infty$ for
 both $k=0$ and $k=\pi$, there exist two intermediate values $k_*$ (one close to $k\to0$ and other close to $k\to\pi$) for which $p_k=1/2$. For these two $k_*$ values, the line of Fisher zeros will indeed cross  the imaginary axis  marking DPTs at two different instants of time for each value of the integer $n$ (Eq.~({\ref{eq_time}})). This is numerically verified and shown in Figs. (\ref{fig:SS2a})
 and  (\ref{fig:SS2b}). {Interestingly, we note that
 corresponding to  each lobe (denoted by $n$), there are two  distinct time instants $t_n^{*}=t_n^{\pm }$ (where Fisher zeros cross
 the imaginary axis) emerging due to  the passage through two QCPs with
 {
 \ba
  t_n^{+}  &=& \frac{(2n+1)\pi}{2\epsilon_{(\pi -k_*)}^f}~~~~~ \nonumber\\
 t_n^{-}&=&\frac{(2n+1)\pi}{\epsilon_{k_*}^f}~~~~~
\label{tns}.
\ea
 The DPT corresponding to the quench across the
 QCP at $h=-1$ occurs at $t_n^{+}$ while the instant $t_n^{-}$ is associated with the QCP at $h=1$
and $t_n^{-} > t_n^{+}$. }
%
%
 (The physical
interpretation of this notation will be clear following the discussion of the next section where a positive (negative) topological charge will be
attributed to a DPT occurring at $t_n^{(+)} (t_n^{(-)}$).) In the case of quenching across a 
 single QCP in Fig.~\ref{fig:SS1a}, we only get zeros corresponding to  $t_n^\ast=t_n^+$.
We can contrast this result  with the case where $h$ is changed abruptly from a large positive to a negative
 value   across both the critical points, one can straightway verify that   
 the lines of Fisher zeros never cross the imaginary axis
 and hence no DPT is observed as predicted in the earlier study by Heyl et al. \ct{heyl13}.

 \begin{figure}
\centering
\subfigure[]
{\includegraphics[width=7cm]{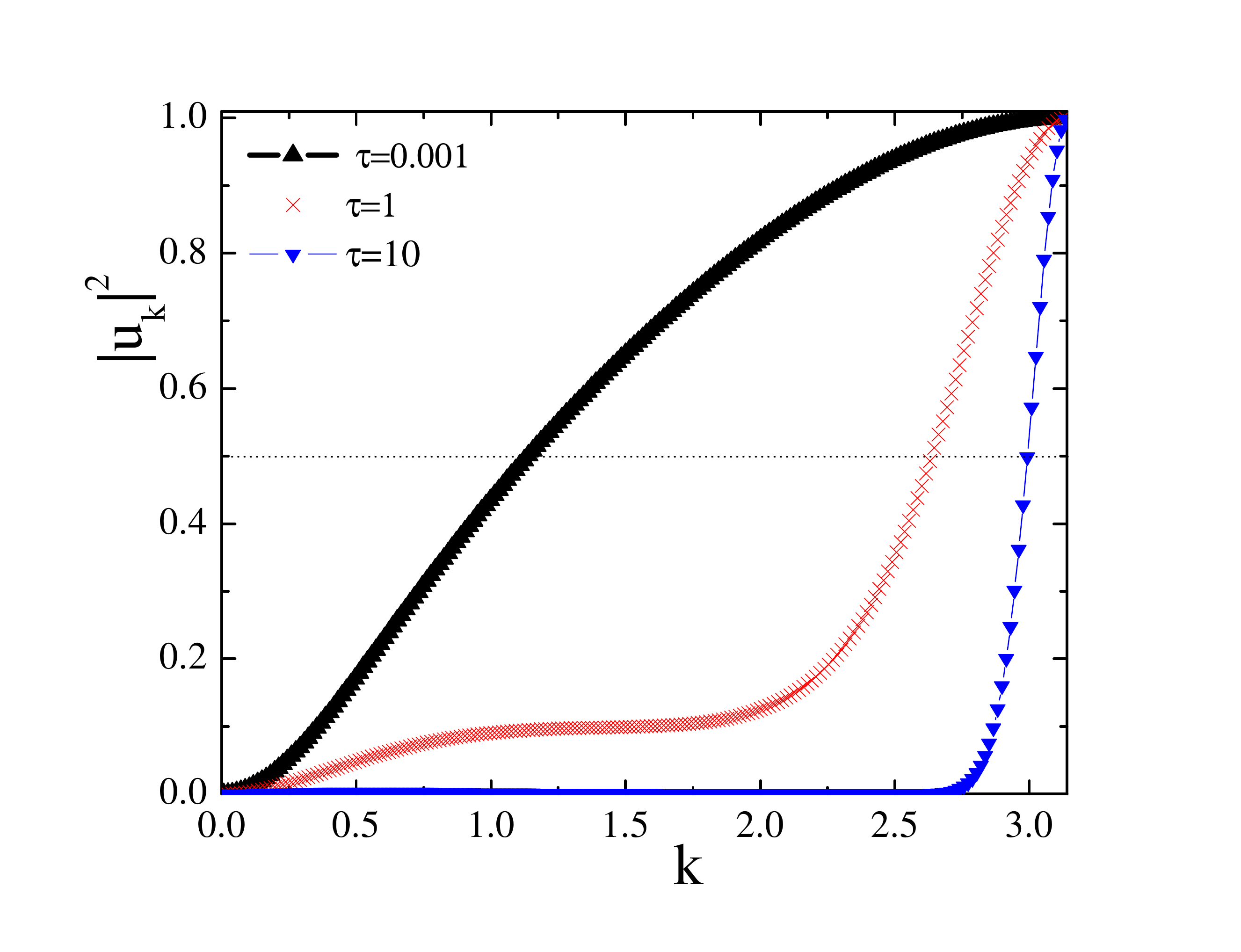}\label{fig:SS_pk_half}}
\subfigure[]
{\includegraphics[width=7cm]{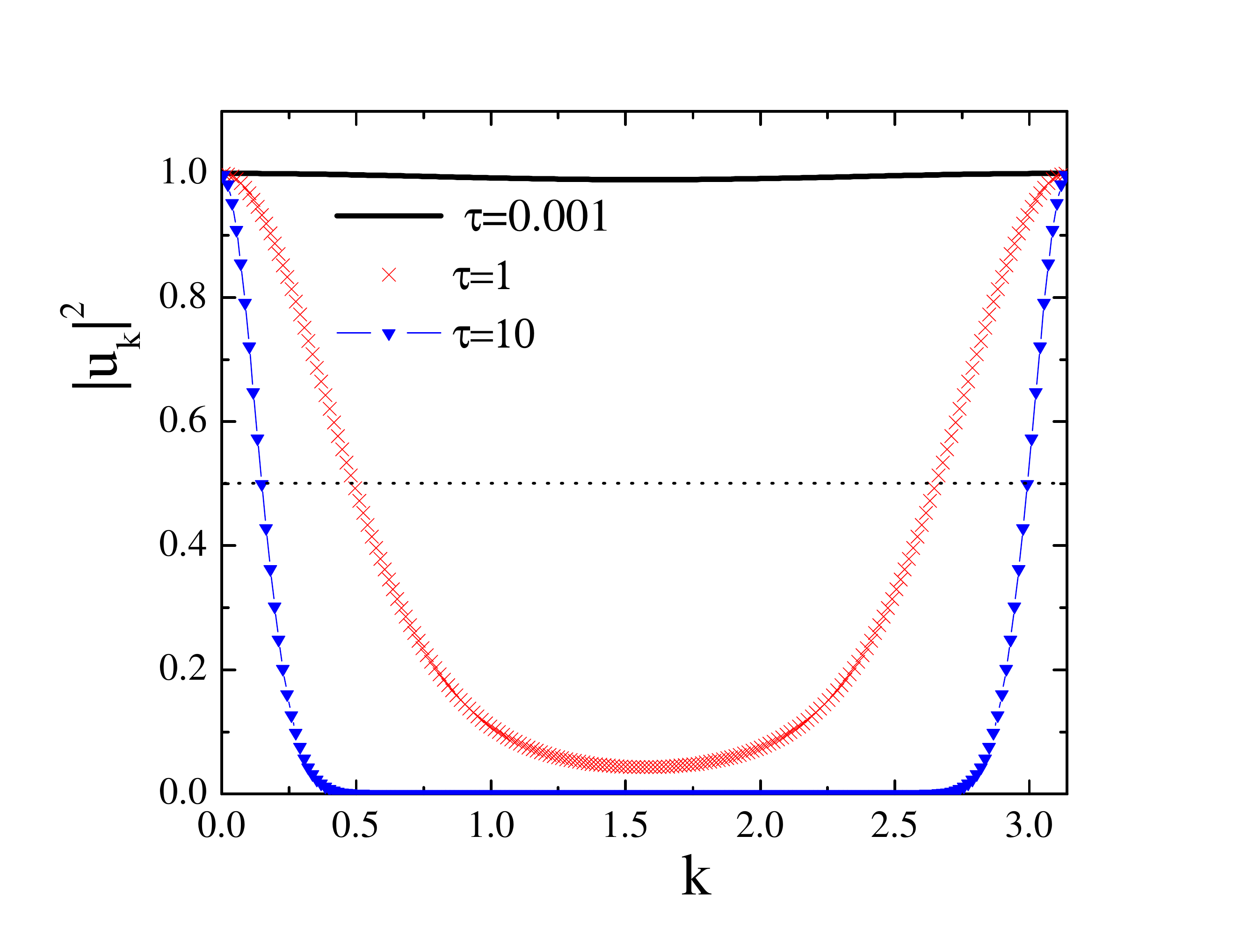}\label{fig:SS_pk_full}}
\caption{(Color online) Numerically obtained values of the adiabatic transition probability (or the probability of the excited state) $p_k = |u_k|^2 = |\langle\psi_{f_k}|2_f^k\rangle|^2$, for quenching across a single QCP (Fig.~(a)) and two QCPs (Fig.~(b)) plotted as a function of $k$ for several values
of $\tau$. The horizontal line denotes $p_k=1/2$. Fig. (a) confirms that the value of $k_*$ depend on $\tau$ and there exists
 a value of  $k_*$ even in the limit $\tau \to 0$. On the other hand, as shown in  Fig.~(b), there are two $k_*$ for each  $\tau$ leading
to two instants of time $t_n^+$ and $t_n^{-}$ where DPTs occur for each lobe denoted by $n$ as shown in Fig.~
\ref{fig:DPT_slow2}. The value $k_*$ depend on 
$\tau$ and in the limit $\tau \to 0$, the situation becomes similar to the sudden quenching case as $|u_k|^2 > 1/2$ for
all values of $k$ and hence no $k_*$ exists.}
\end{figure}
 
 {The slow quenching scheme analyzed here brings in another time scale in the problem, namely, the
 inverse quenching rate $\tau$. A pertinent question at this point would be how does the scenario depicted in
 Figs.~\ref{fig:DPT_slow1} and \ref{fig:DPT_slow2} gets altered when $\tau$ is varied. Let us first consider the case of crossing
 a single QCP at $h=-1$ ( Fig.~\ref{fig:DPT_slow1}), where gap vanishes for the critical mode $k=\pi$;  expanding around this mode, one can
 rewrite the reduced $2\times 2$ Hamiltonian 
 \begin{equation}
 H_{k}= 
 \left(
 \begin{array}{cc}
-h -1 + \frac {(\pi-k)^2}{2} & -i  (\pi -k)  \\
i (\pi -k) &  h +1-\frac{(\pi-k)^2}{2}  \\
 \end{array}
 \right).
 \label{ham_2by2_small_k}
 \end{equation} 
Analyzing the Hamiltonian \eqref{ham_2by2_small_k}, we shall now probe the non-adiabatic transition probability (i.e., the probability of the excited state $|2_k^f\rangle$)  given by $p_{k} = |u_{k}|^2$, at the end of the quenching process; it is obvious  that the mode $k=\pi$ is temporally frozen as the off-diagonal
term vanishes, and hence  $|u_{k=\pi}|^2 = 1$. On the other hand, modes away from $k=\pi$ evolve adiabatically
and hence $|u_{k \ll \pi}|^2=0$. From the continuity argument, one then immediately expects a value of $k=k_*$ for which
$p_{k=k_*} = |u_{k=k_*}|^2 =1/2$ which is essential for observing a DPT. In Fig.~\ref{fig:SS_pk_half}, we plot numerically obtained
$|u_k|^2$ as a function of $k$; indeed an analytical expression for the same can be derived using the Landau-Zener (LZ)
transition formula for non-adiabatic transition probability  \ct{landau,vitanov96, sei,grandi10} which can be approximated as $|u_{k \sim \pi}|^2=\exp(-\pi (\pi-k)^2 \tau)$. From
these observations, it is straightforward to conclude that  it is the value of $k_*$ and hence, those of $t_n^*$, related through
Eq.~\eqref{eq_time}, which get
modified when $\tau$ is changed.  On the other hand,  we would like to emphasize that the behavior of Fisher zeros and non-analyticities in $I(t)$ do not change qualitatively; the non-analyticities only occur at different instants of time depending upon the
value of $\tau$. This claim will be further illustrated  in Fig.~\ref{fig:SS3b}.  The existence of DPTs for $\tau\to0$ is expected since for a large amplitude sudden quench across a single QCP a $k_*$ will always  be present as shown in Fig.~\ref{fig:SS_pk_half} \ct{heyl13}.}



{We now proceed to analyze the situation  presented in Fig.~\ref{fig:DPT_slow2};
in this case the system crosses both the critical points at $h=-1$ (with critical mode $k=\pi$) and $h=1$ (with critical
mode $k=0$). As a results both the modes $k=0$ and $k=\pi$ are temporally frozen leading to $|u_{k=0}|^2 =|u_{k=\pi}|^2 =1$
(see Fig.~\ref{fig:SS_pk_full}); the corresponding LZ  formula is $p_k= |u_k|^2=e^{-\pi\tau \sin^2{k} }$. Therefore, for each value of $\tau$, we 
find two values of $k_*$ (with $p_{k=k_*}=1/2$)  yielding two real instants given by $t_n^+$ and $t_n^{-}$ as defined before. We therefore again conclude
that only the instants of real time at which DPTs occur will depend on the inverse rate $\tau$. 
Referring to Eq.~\eqref{eq_time}, we
note that $t_n^{+}$ or $t_n^{-}$ depends on $\epsilon_{k_*}^f=2\sqrt{(h_f -\cos k_*)^2+\sin^2k_*}$, where $k_*$ assumes
two values (close to $k=0$ and $k=\pi$).
However, in the limit of small $\tau (\to 0)$, (i.e., in the sudden limit) as presented  in Fig.~\ref{fig:SS_pk_full}, one does not find
 a value of $k_*$ and no DPT is expected. Furthermore, it can be shown in
  the following straightforward manner that if quenching is fast (i.e., $\tau <  \ln 2/\pi$), the DPTs disappear for 
 passage across two critical points:  analyzing  the profile of $p_k= |u_k|^2=e^{-\pi\tau \sin^2{k} }$,
we note that  the minimum value of $p_k$ (that occurs at $k=\pi/2$) should  at least be less  than $1/2$, for DPTs to be present. This implies that   $p_{k\to\pi/2}=p_k^{\rm min}=e^{-\pi \tau_c}=1/2$; here $\tau_c$ refers to a critical $\tau$ and   DPTs exist for $\tau > \tau_c$; otherwise, the situation is identical to the sudden quenching across two QCPs where DPTs get wiped out. }

\section{Topological Aspects:}
\label{sec_topology}
Topological properties of DPTs  are related to topological structure of equilibrium phase diagram, which for the present model in turn, is determined by the sign of the quantity $(h+\cos k)_{k=0}/
(h+\cos k)_{k=\pi}=(h+1)/(h-1)$; 
clearly it is negative in the topological (ferromagnetic phase with $|h|<1$) while positive in the non-topological (paramagnetic) phase, $|h|>1$.
We now proceed to investigate the behavior of the DTOP; to define it in the present context, we refer  to the LO:
\be { L}_k= \left( |v_k|^2 + |u_k|^2 \exp(-2i \epsilon_k^f t) \right).
\label{eq_topology1}
\ee

      {Let us first consider the situation of quenching from the non-topological (paramagnetic) to the
topological (ferromagnetic) phase across a single QCP (take for example, the situation illustrated in the Fig.~\ref{fig:SS1a}). In  a spirit similar to the Ref. [\onlinecite{budich15}], we can also express the LO as  $L_k = |r_k| \exp(i {\phi_k})$. 
 In the present case of slow quenching, 
 \be
 \phi_k = \tan^{-1} \left(\frac{-|u_k|^2 \sin(2\epsilon_k^f t)}{|v_k|^2 + |u_k|^2 \cos (2\epsilon_k^f t)} \right),
 \label{eq_phi1}
 \ee
 and the corresponding dynamical phase  $\phi_k^{\rm dyn} = -\int_0^t ds \langle \psi_{f_k}(s)|H_f|\psi_{f_k}(s)\rangle = -2 |u_k|^2 \epsilon_k^f t$ with $|\psi_{f_k}\rangle = v_k |1_k^f\rangle + u_k e^{-2i\epsilon_k^f s} |2_k^f\rangle$.
Let us now define the  geometric phase as 
 $\phi_k^G =\phi_k -\phi_k^{\rm dyn}$, so that
   \begin{widetext}
 \be \phi_k^G =   \tan^{-1} \left(\frac{-|u_k|^2 \sin(2\epsilon_k^f t)}{|v_k|^2 + |u_k|^2 \cos (2\epsilon_k^f t)} \right)  + 2 |u_k|^2 \epsilon_k^f t .\ee
 \end{widetext}
 {Note that for the mode $k=0$, we have $|v_{k=0}|^2=1 $, and the excitation probability $|u_{k=0}|^2=0$; on the other hand, for the critical
 mode $k=\pi$, $|v_{k=\pi}|^2=0 $ and $|u_{k=\pi}|^2=1$.  The geometric phase therefore satisfies the periodicity relation:
\be
\phi^G_\pi-\phi^G_0=0 \mod 2\pi
\ee
As the lattice momentum goes from $0$ to $\pi$, $\phi^G_k$ goes from $-\pi$ to $\pi$ thus completing a full circle (See Fig. ~\ref{fig_phi_G}).
Focusing  on the  mode $k_{*}$, we find that
\be 
\phi_k^G|_{k_*} =   \tan^{-1} \left(-\tan (\epsilon_{k_*}^f t) \right)  + 2 |u_{k_*}|^2 \epsilon_{k_*}^f t, 
\ee
with $ |u_{k_*}|^2 =1/2$; this shows that  $\phi_k^G|_{k_*}$ is fixed to zero or $\pi$   for all values of $t$ except at DPTs given by the condition $ \epsilon_{k_*}^f t_n^*= (n+1/2)\pi$ where $\phi_k(t)$ is ill-defined; $\phi_k^G$ alternates between $0$
 and $\pi$ between two DPTs  at $k=k_{*}$.
}
 

We study the behavior of DTOP or
  the winding number \be
  \nu_D = \frac 1{2\pi} \oint_0^{\pi} \frac {\partial \phi_k^G}{\partial k}, \label{eq_winding}
\ee
  close to DPTs (discussed in the previous section) focusing on the quenching across a single QCP first.    
   In the context of sudden quenches, the DTOP has been found to  appropriately characterize a DPT \ct{budich15}. 
   We address the question whether the same is true in
   the case of slow quenches and show that it is indeed the case.
   
{ As we are integrating in Eq.~\ref{eq_winding} the full derivative of a periodic function (modulo $2 \pi$), the integral remains constant unless there is some discontinuity in the phase, which should be manifested in the $\delta$-function type contribution to the derivative of the geometric phase. We can anticipate that such discontinuities develop only at DPT therefore it suffices to explore}
  the  quantity ${\partial \phi^G_k}/{\partial k} $ for the  momentum $k_*$:
  
  \be
 \frac{\partial \phi^G_k}{\partial k}\bigr|_{k_*}  =  2 \tan (\epsilon_{k_*}^f t) \frac{\partial |v_k|^2}{\partial k}\bigr|_{k_*} + 2  \frac{\partial |u_k|^2}{\partial k}\bigr|_{k_*} (\epsilon_{k_*}^f t).
 \label{eq_partial_phi}
  \ee
%
While the first term diverges at every DPT, the second term provides a linearly varying  contribution with a slope determined
by the sign of its coefficient $(\partial_k |u_k|^2)|_{k_*}$. (In fact,
it can be shown that the term $\oint \partial_k \phi_k$ varies symmetrically  between positive and negative
values in the intermediate time between two DPTs while $\oint \partial_k \phi^{\rm dyn}_k$ increases or decreases
monotonically with time depending on the sign of $(\partial_k |u_k|^2)|_{k_*}$).
It is therefore clear that the sign of the change in $\nu_D$ at a DPT will be determined by the sign of the coefficient of the linearly increasing term  $(\partial_k |u_k|^2)|_{k_*}(=-(\partial_k |v_k|^2)|_{k_*})$.
(This has been loosely connected to an index theorem in ref. [\onlinecite{budich15}].) 
 Positive sign of $(\partial_k |u_k|^2)|_{k_*}$ corresponds
to a jump $\Delta \nu_D=+1$ while  negative sign  yields  $\Delta \nu_D=-1$ at every DPT.
 }   
   
 \begin{figure}
 \begin{center}
  {\includegraphics[width=6cm,angle=-90]{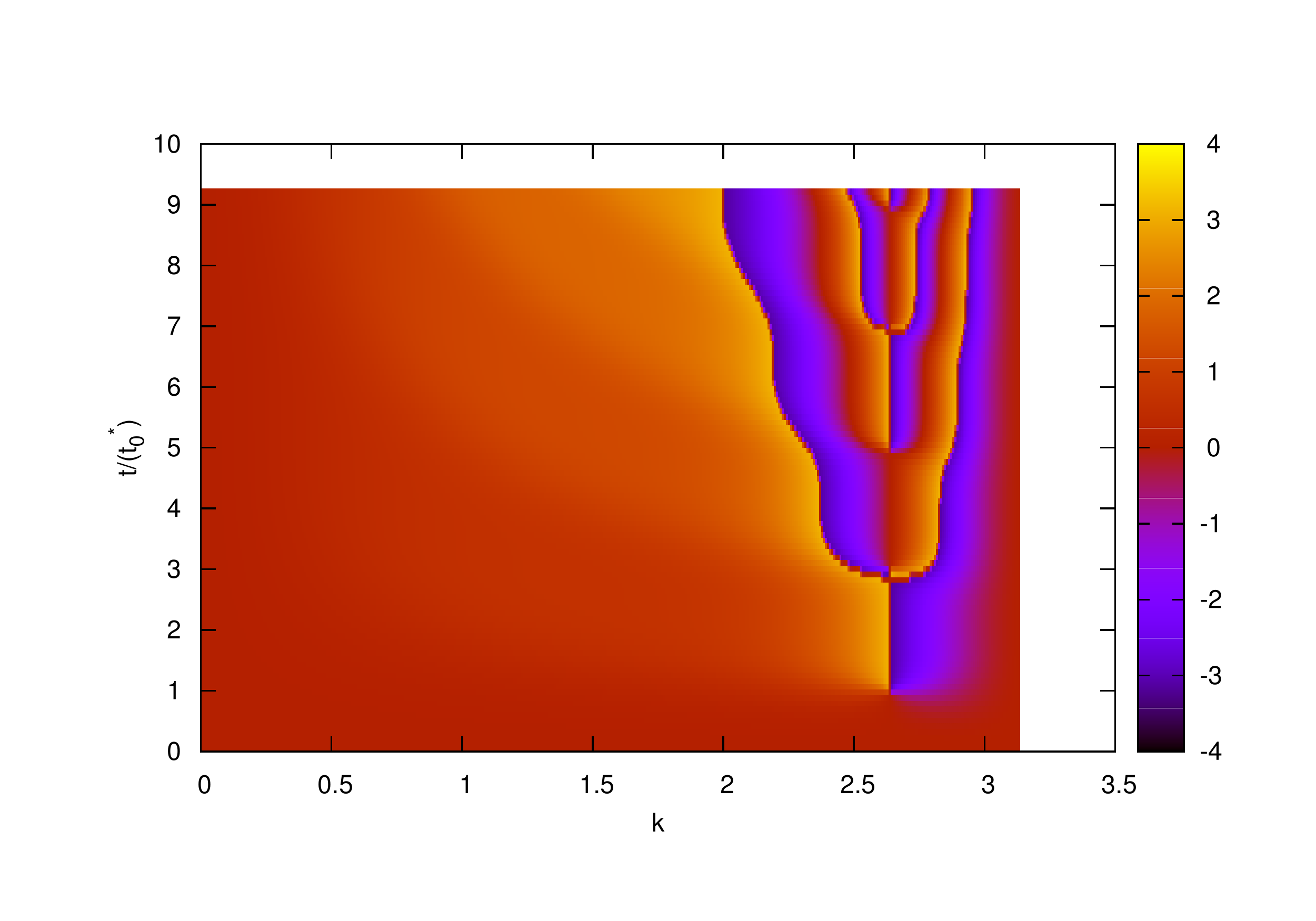}}
  \end{center}
  \caption{(Color online) Variation of $\phi_k^G$ as a function of $k$ and $t/t_0^*$ (where $t_0^* = \pi/2\epsilon_{k_*}^f$) following a slow quench $h({\tilde t})={\tilde t}/\tau$ {(with $\tau=1$)} from the initial value of the field $h_i= -10$ to the final value $h_f =0.5$ across the QCP at $h=-1$. As the lattice momentum $k$ goes from $0$ to $\pi$, $\phi_k^G$
 varies from $-\pi$ to $+\pi$.  It is to be noted that  between $t_0^*$ and $3t_0^*$ there is a single winding
resulting in $\nu_D =1$ while between  $3t_0^*$ and $5t_0^*$ two full winding and hence $\nu_D =2$. }
\label{fig_phi_G}
\end{figure}

 {Fig.~\ref{fig:SS3} shows the variation of $\nu_D$ as a function of time following a quench from large negative $h$ to the topological
phase (e.g., $h_f=0.5$); $\nu_D(t)$ increases in a step-like fashion at every DPT with $\Delta \nu_D=1$. Let us compare  with Fig.~2 of Ref. [\onlinecite{budich15}], where a sudden quench from the topological phase
to the
 trivial phase of a $p$-wave superconducting chain was studied.  As discussed above,  the sign of $\Delta \nu_D$  at a DPT is given in
terms of the sign of $(\partial_k |u_k|^2|)|_{k_*}$; this is negative in the situation studied in Ref. [\onlinecite{budich15}]
where $ |u_{k=0}|^2 =1$ and $|u_{k=\pi}|^2 =0$. On the contrary,  for a slow quenching  of the transverse Ising chain 
starting from  large negative $h$ across the QCP at $h=-1$ (with the critical
mode $k=\pi$),
we find the value of $\nu_D$ increases at every
DPT. 
Here,
 $|u_{k=\pi}|^2=1$ and $|u_{k=0}|^2=0$ ; 
from continuity one expects a $k_{*}$ for which  $|u_{k=k_*}|^2= |v_{k=k_*}|^2=1/2$  and hence,  $\Delta \nu_D(t_c) = 
 {\rm sgn} (\partial_k |u_k|^2|)|_{k_*}$ is  positive. 
To establish that this is indeed the case, we reverse the
 quenching scheme to {$h({\tilde t})={\tilde t}/\tau$ with $\tilde {t}$ starting from a large positive  value  and ending in the topological phase.} The spin chain
 then crosses the QCP at $h=1$ where the gap vanishes for the mode $k=0$ and hence $|u_{k=\pi}|^2=0$ and $|u_{k=0}|^2=1$,
leading to  a negative  $\Delta \nu_D(t_n^*) = 
 {\rm sgn} (\partial_k |u_k|^2|)|_{k_*}$;  
 one can straightway verify numerically that  $\Delta \nu_D$ is indeed negative in this case
 (see Fig.~\ref{fig:SS3A}). Furthermore, one can investigate the variation $\partial \nu_D/\partial t$ as function of time: $\nu_D$ is independent of time always
except at  DPTs ($t_n^*$) when  there is a diverging
($\delta$-function) peak.}

      {Equipped with above observations, we can now interpret Eq.~(\ref{eq_winding}) as a  generalized Gauss's law;
as the system evolves in time following a quench across a single QCP, one can assume that a positive topological  charge enters the system at every $t_n^*$. On the other hand, if the system is initially
chosen to be in  the non-topological phase with a large positive $h$ (i.e., the situation depicted in Fig.~\ref{fig:SS3A}), we can infer  that a negative change enters the system at every DPT.  }
 
  \begin{figure}
  \subfigure[]
  {\includegraphics[width=8cm]{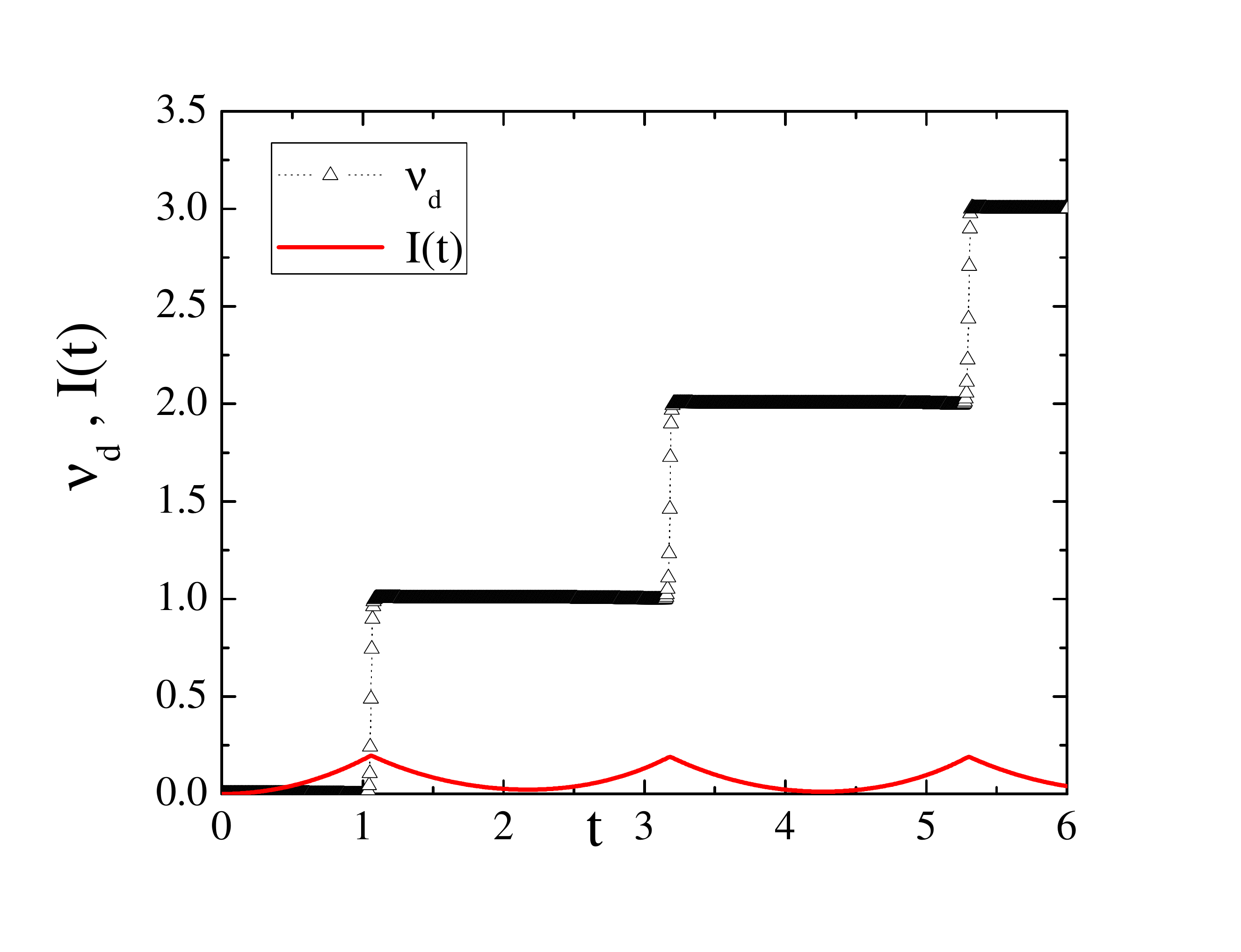}}
  \subfigure[]
  {\includegraphics[width=9cm]{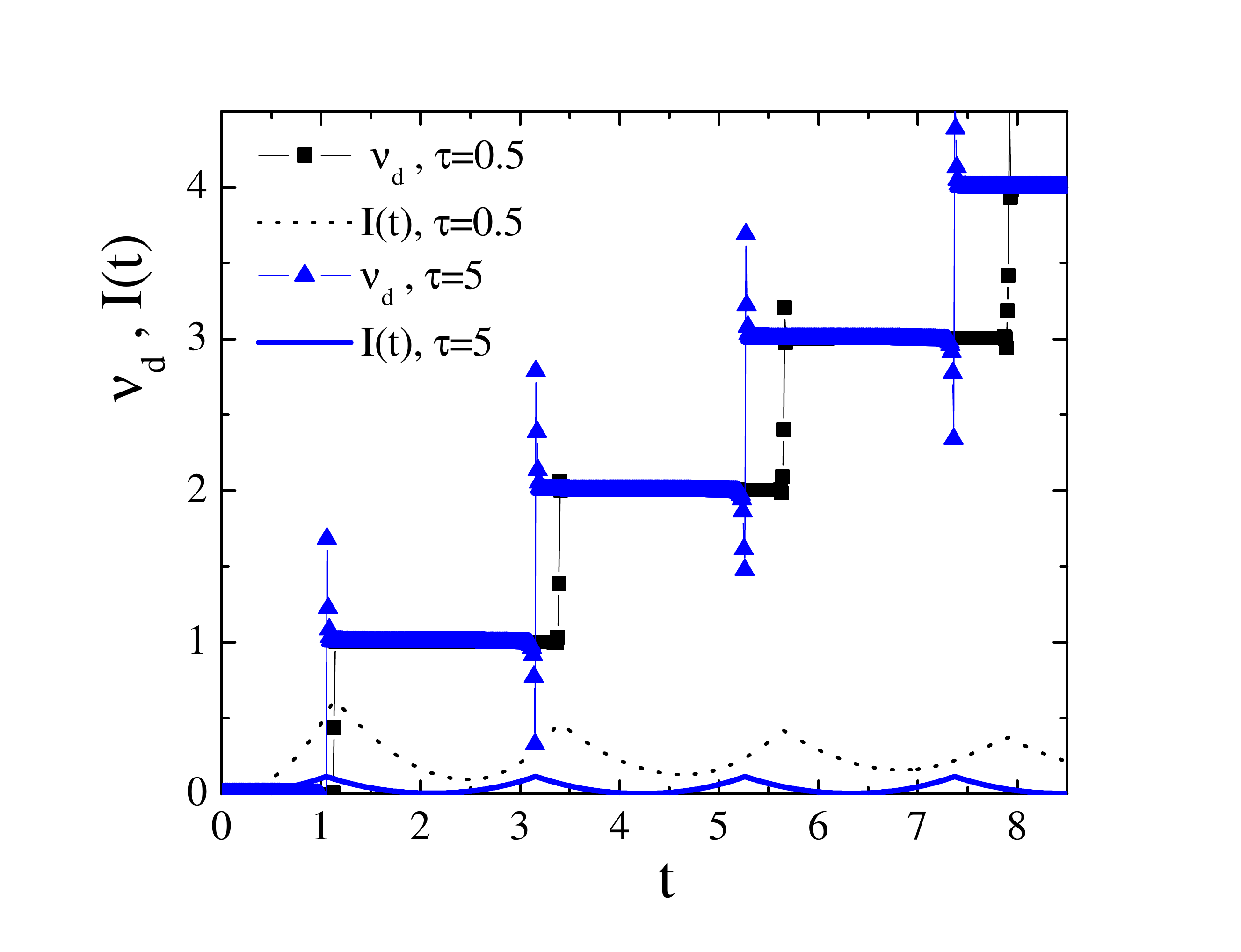}\label{fig:SS3b}}
  \caption{(Color online) (a) $\nu_D$  along with the rate
  function $I(t)$ are plotted as a function of time $t$ following a slow quench $h ({\tilde t})= {\tilde t}/\tau$ ({with $\tau=1$}) from $h_i =-10$ to $h_f =0.5$ so that the system
  crosses  the QCP at $h=-1$. We find an increases in $\nu_D$ by a factor of
  unity whenever there is a DPT. (b) {Using two different values of  $\tau$ ($0.5$ and $5$), we show that the time instants at which DPTs  
  occur  depend on $\tau$. However, $\nu_D(t)$ indeed shows a jump of unit magnitude
  at every DPT for all $\tau$ as explained in the text.}  }
\label{fig:SS3}
\end{figure}

\begin{figure}
  {\includegraphics[width=8cm]{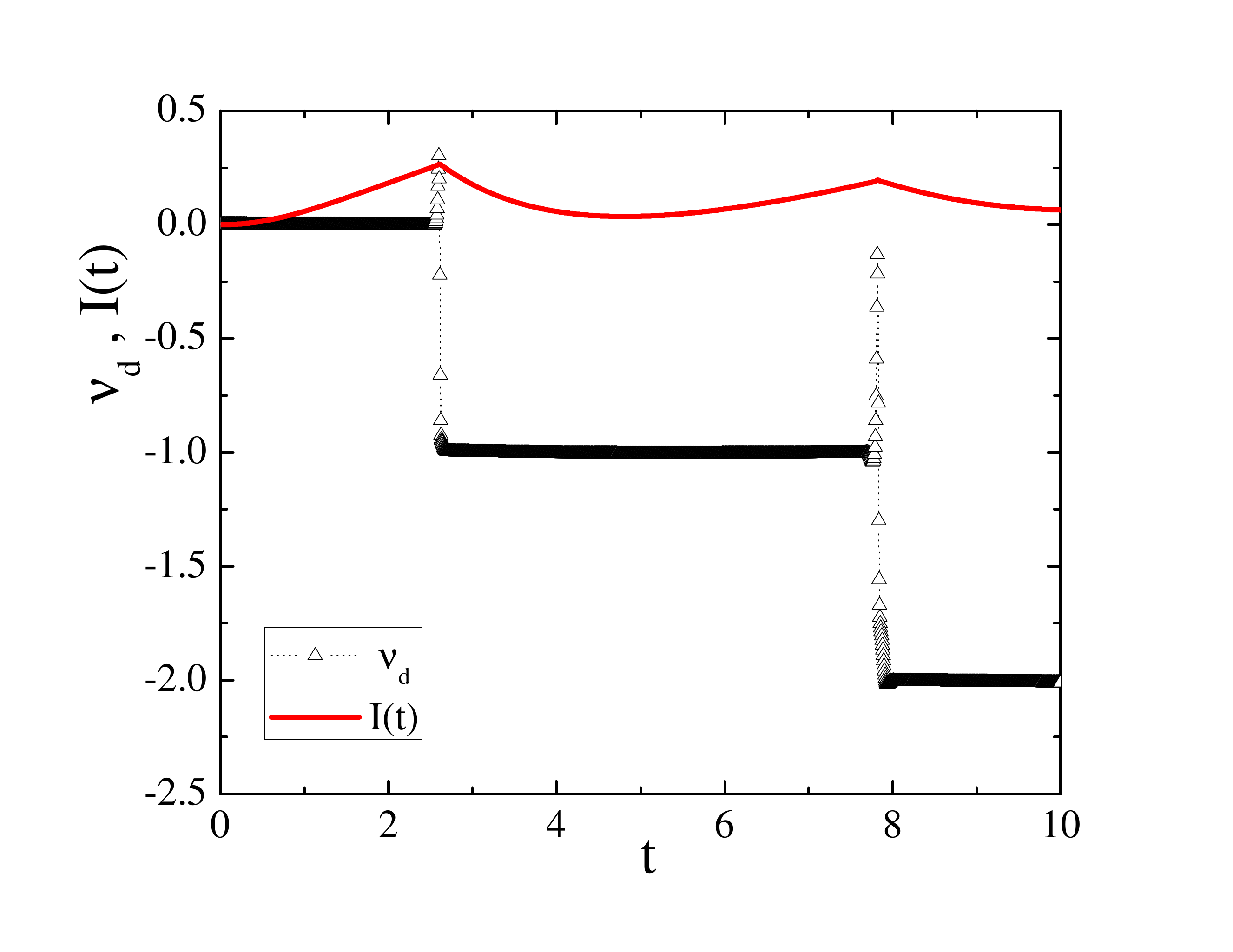}}
  \caption{(Color online) This shows that when the system quenched across the QCP at $h=1$ to the topological phase ($h_f=0.5$) starting
  from a large positive $h$ ($h_i=10$), ${\rm sgn} (\partial_k |u_k|^2|_{k_*})$, and hence $\Delta \nu_D$ is indeed negative.  }
\label{fig:SS3A}
\end{figure}

 \begin{figure}
  {\includegraphics[width=8cm]{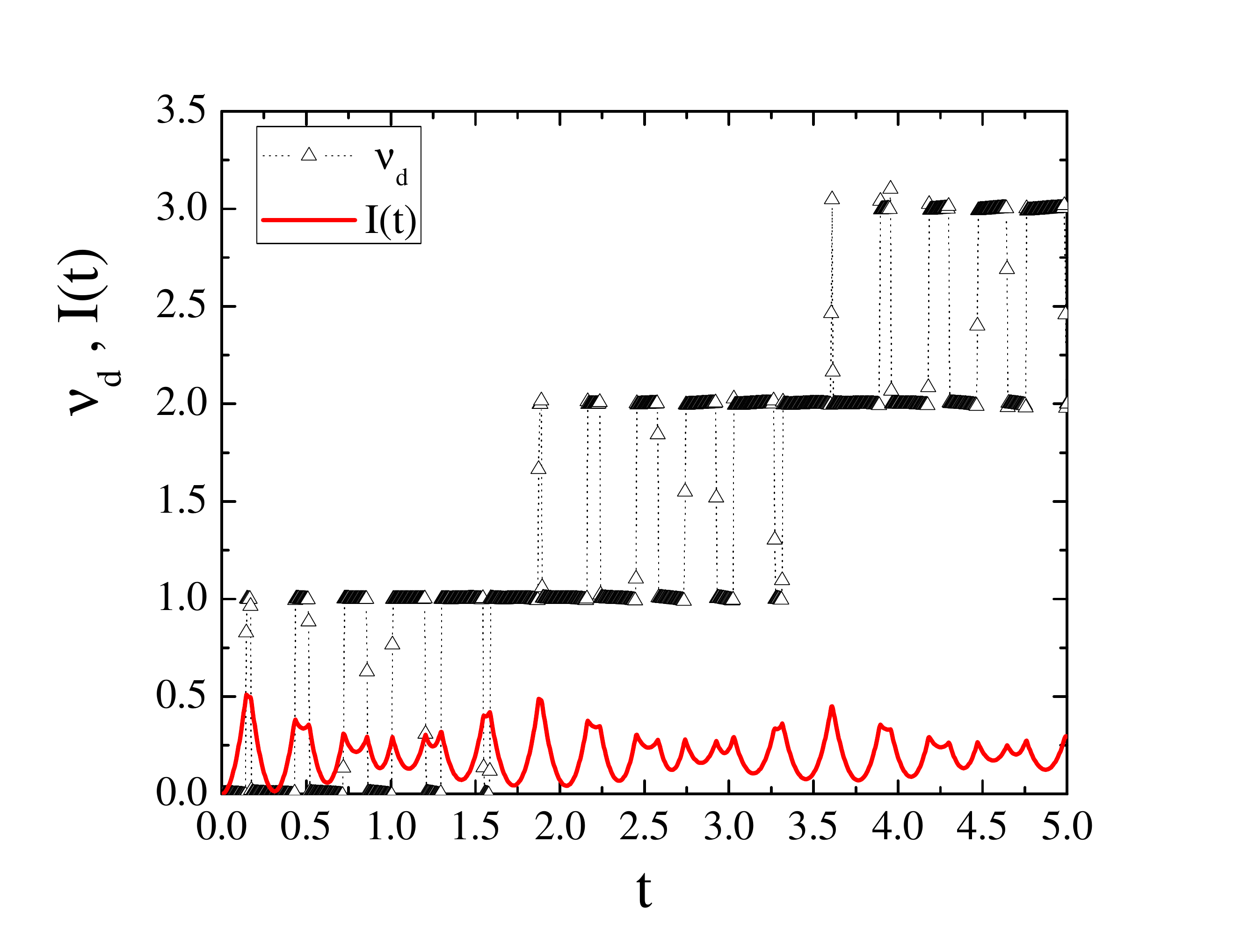}}
  \caption{ (Color online) $\nu_D$ along with the rate
  function $I(t)$ are plotted as a function of time $t$ following a slow quench $h ({\tilde t})= {\tilde t}/\tau$ from $h_i =-10$ to $h_f =10$, 
   so that the system
  crosses both the QCPs at $h=-1$ and $h=+1$.  We find that $\nu_D$ oscillates
  between $0$ and $1$, i.e., $\nu_D = \pm 1$ at successive DPTs. However there is a   jump in $\nu_D$ by unity for
  some values of $t$       {when there are two successive DPTs with $\Delta \nu_D=1$ (or when a positive
  charge crosses a negative charge as shown in Fig.~\ref{fig_SS5})}; between two jumps  the alternating pattern persists. }
  \label{fig:SS4}
  \end{figure}

  

      {Let us now address the question how does the  above  situation gets altered} when
       the field is quenched from a large negative to a large positive value so that
the system is swept across both the QCPs at $h=\pm 1$, from one topologically trivial phase to the other. In this case as well $\nu_D$ exhibits a discrete change at every DPT, however,
there exist  further intricacies as  presented in Fig.~\ref{fig:SS4}.        {To analyze this remarkable  behavior, we plot the flow of Fisher zeros following a double quench as shown in the
Fig.~\ref{fig_SS5}. We note that corresponding to each lobe  denoted by $n$, there are two DPTs occurring at
two instants of times $t_n^{(+)}$ and $t_n^{(-)}$ with  $t_n^{(-)} > t_n^{(+)}$.  From Fig.~(\ref{fig:SS4}), it is evident that at $t_n^{(+)}$, $\Delta \nu_D =1$ (in other words, a positive charge enters the system as shown
in Fig.~\ref{fig_SS5}); on the
other hand, the DPT at  $t_n^{(-)}$ is associated with  $\Delta \nu_D =-1$ or a negative topological  charge. This makes $\nu_D$ oscillate between $0$ and $1$. This pattern follows
up to 5th lobe and there is a discontinuity at the 6th lobe: from Fig.~\ref{fig_SS5}, we find  that $t_6^{-} > t_7^{+}$. Therefore,
 we have two successive DPTs  both with $\Delta \nu_D =+1$ (i.e., with identical topological charges) and consequently, $\nu_D$ jumps from $1$ to $2$. In general, this type of change would first happen
when $t_n^{(-)} > t_{n+1}^{(+)}$, i.e., second DPT (with a negative topological charge) corresponding to the lobe $n$ 
happens at a later (real) time 
compared to the first DPT (with a positive charge) associated with $(n+1)$-th lobe. }

Finally, let us discuss the issue concerning the analytic continuation of the boundary partition function to the real time
axis. Comparing Figs.~\ref{fig:SS1a} and \ref{fig:SS2a} we see a very different topology pattern of the Fisher zeros for the quenches across one and two critical points. In the former case Fisher zeros slice the complex time space into disconnected regions. As discussed in Ref.~[\onlinecite{heyl13}] this prevents analytic continuation of a boundary partition function defined as $Z_\psi(z)=\langle \psi| \exp[-z H]|\psi\rangle$ into the real time-axis beyond the critical time $t^\ast$. On the other hand, for the quench across two critical points the time domain is split into  alternating intervals. The analytic continuation of $Z(z)$ is only possible to the intervals reachable without crossing Fisher zeros, {i.e. $0<t<t_0^+$, $t_0^{-}<t<t_1^+$} and so on. As we discussed above such intervals where analytic continuation is possible correspond to the sectors with zero topological charge. As it is clear from Figs.~\ref{fig:SS4} and \ref{fig_SS5}  such piecewise analytic continuation is possible for finite time until the zero with positive topological change from the $n+1$-st branch of Fisher zeros crosses the zero with negative topological charge  from the $n$-th branch, i.e. $t^{+}_{n+1}<t^-_n$. Beyond this point no analytic continuation of the boundary partition function is possible.

{Finally, we address the question how does this topological structure depend on $\tau$; as argued
in the previous section, $\tau$ determines the time instants at which DPTs occur. Hence there will not be any qualitative
change in the topological pattern, i.e.,  $\Delta \nu_D =1$ ( or $-1$) at every DPT as illustrated in Fig.~\ref{fig:SS3b}. However, in the case of crossing across
two QCPs  the values of $n$ for which $t^{+}_{n+1}<t^-_n$ and hence  we have two successive DPTs with $\Delta \nu_D=1$
may depend on $\tau$.
 }

\section{Concluding comments}

 We have studied the slow quenching dynamics of a transverse Ising chain across its QCPs and analyzed the Fisher
zeros and hence DPTs which are reflected in the non-analyticities of the rate function. We emphasize that although the non-analyticities 
in the dynamical free energy leading to periodic occurrences of DPTs have been reported in the context of slow quenching of 
the Hamiltonian (\ref{eq_hamil_txy}) \cite{pollmann10}, they have not been connected to the Fisher zeros. Furthermore, we
show that these DPTs survive even when the system is quenched across both the QCPs of the model when the line of Fisher zeros
crosses the imaginary axis for two characteristic momenta values. This is in sharp contrast with the sudden quenching case when
an abrupt passage across two QCPs wipes out the non-analyticities. In fact, for a sufficiently slow quenching, two QCPs of
the model appear to be independent of each other.
       {Secondly, in this case, it is essential to cross the
QCP to observe the DPTs, otherwise the transition probability $p_k$ is always less than $1/2$ for all values of $k$ and hence  a DPT can not occur}. 

We have also analyzed the behavior of the DTOP (i.e., $\nu_D(t)$). The slow quenching scheme discussed in this paper, 
allows to study both cases:
the quenching across a single QCP (i.e., from a non-topological phase to the topological phase) or two QCPs (i.e., from
one non-topological phase to the other) depending upon the final value of the transverse field. In the former case, the DTOP increases (or
decreases depending upon the direction of the quenching) step-wise by  unity at those instants of time where DPTs occur.
 {This can be interpreted as charges of the same sign entering the system at every DPT.}
On the contrary, in the latter case the DTOP oscillates between 1 and 0;       {this implies that a DPT with a positive topological charge
$\Delta \nu_D=1$ at real time $t_n^{+}$ is followed by a  DPT with a negative topological charge  $\Delta \nu_D=-1$ at $t_n^-$. Furthermore there are sharp increases by
a factor of unity whenever there is a crossing of Fisher zeros belonging to neighboring lobes i.e., $t_n^{-} > t_{n+1}^+$, there are
two successive DPTs with topological charges of same signs}; this is indeed remarkable.

\begin{figure*}[ht]
\includegraphics[height=11.2cm,width=17.9cm]{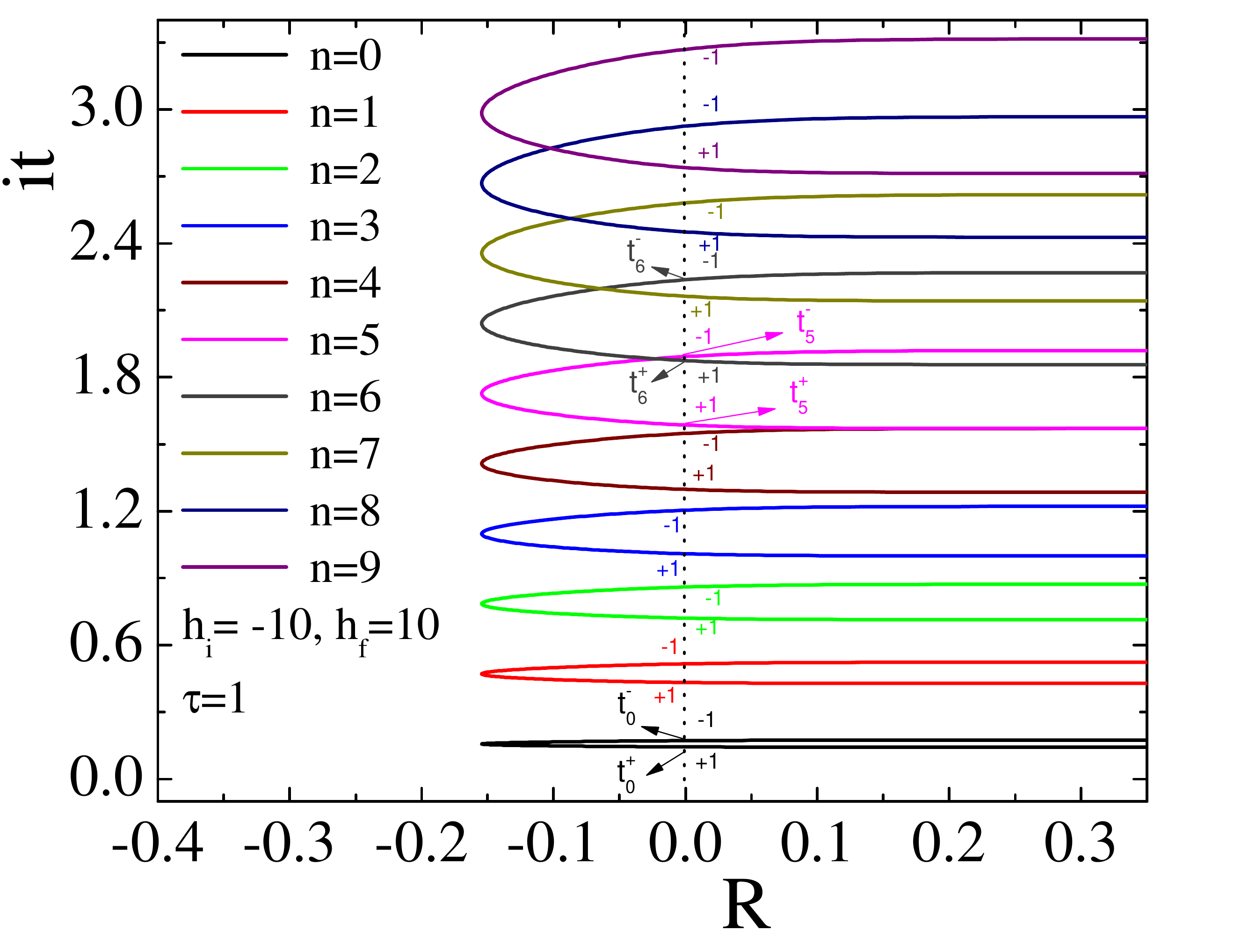}
 \caption{(Color online) The flow of  Fisher zeros following quenching  from $h_i=-10$ to $h_f =10$ across two QCPs {with
 the protocol $h({\tilde t}) ={\tilde t}/\tau$ with $\tau=1$}. Each lobe is characterized by the
 integer $n$ which increases
 from bottom to top with the lowest lobe corresponding to $n=0$.       {There exist two DPTs associated with each $n$; the
 DPT at $t_n^+$ is associated with a positive topological charge ($+1$) while $t_n^-$ denotes a DPT with a 
 negative topological charge ($-1$). This explains the oscillation of $\nu_D$ between $0$ and $1$. However, we
 note  that $t_6^+ < t_5^-$ and hence there are two successive DPTs both with positive topological charge occurring at $t_5^+$
 and $t_6^+$, respectively; this leads to 
 a jump in $\nu_D(t)$ by a factor of unity (as shown in Fig.~\ref{fig:SS4}) following which the oscillating pattern  persists.}
 }
 \label{fig_SS5}
\end{figure*}

\begin{acknowledgments}
We acknowledge Jun-ichi Inoue  for discussions and Sei Suzuki for extensive discussions, critical comments on the manuscript and collaboration in related works. Special thanks to Utso Bhattacharya for his critical comments. SS acknowledges
CSIR, India and also DST, India, and AD  and UD acknowledge DST, India,  for financial support. AD and SS acknowledge 
Abdus Salam ICTP for hospitality where the initial part of the work was done. A.P. was supported by AFOSR FA9550-13-1-0039, NSF DMR-1506340, ARO W911NF1410540
\end{acknowledgments}

\vspace{-\baselineskip}

\end{document}